\documentclass[fp,twocolumn]{jpsj3}

\title{Flat Surface State with Octupole Moment in an $e_g$ Orbital System\\
  on a Simple Cubic Lattice}

\author{Katsunori Kubo}
\inst{Advanced Science Research Center,
  Japan Atomic Energy Agency, Tokai, Ibaraki 319-1195, Japan}

\date{\today}

\abst{
A tight-binding model for $e_g$ orbitals on a simple cubic lattice
with finite thickness is investigated.
The hopping integrals for nearest-neighboring sites are considered.
We examine the electronic band structures
for systems with (001), (110), and (111) surfaces.
Electronic states well localized around the surfaces are found
for the (110) and (111) surfaces.
In particular, the surface state is flat and extends
in the entire Brillouin zone for the (111) surface,
provided the bulk band projected onto the surface Brillouin zone is gapped.
We also find that these surface states
possess octupole moments in both the (110) and (111) surface cases.
}

\begin{document}
\maketitle

\section{Introduction}
Topological semimetals are distinguished by
the topological protection of band degeneracies,
which can occur at discrete points, known as Dirac or Weyl points~\cite{Wallace1947, Novoselov2005, Murakami2007, Wan2011},
or along continuous lines, referred to as topological nodal lines~\cite{Heikkila2011, Burkov2011, Weng2015}.
These materials have attracted considerable attention
due to their unique and fascinating physical properties.

Achieving such topological features requires at least two electronic bands.
In systems with two atomic sites per unit cell,
such as tight-binding models with nearest-neighbor hopping
on honeycomb and diamond lattices,
this condition is naturally met,
leading to the presence of Dirac points~\cite{Wallace1947, Novoselov2005}
and nodal lines~\cite{Chadi1975}, respectively.
Another approach involves leveraging the spin degrees of freedom.
For example, the Rashba spin-orbit coupling~\cite{Bychkov1984}
can lift the spin degeneracy in electronic bands,
except at specific $\mib{k}$ points
where time-reversal symmetry ensures degeneracy,
giving rise to Weyl points~\cite{Kubo2024JPSJ}.

Another possibility is to use the orbital degrees of freedom.
We examine the $e_g$ orbitals of $d$ electrons
as a prototypical example of two-orbital systems.
Indeed, Dirac points have been investigated
in two-dimensional $e_g$ orbital systems:
$e_g$ orbital models~\cite{Bishop2016, Kubo2024PRB},
a cuprate superconductor~\cite{Horio2018},
and a LaAlO$_2$/LaNiO$_3$/LaAlO$_3$ quantum well~\cite{Tao2018}.
In particular, in our previous study~\cite{Kubo2024PRB},
we found edge states in an $e_g$ orbital model on a square lattice,
similar to the single-orbital honeycomb lattice model~\cite{Fujita1996}.
The edge state in the $e_g$ orbital model possesses an octupole moment.

Although this $e_g$ orbital model is simple,
considering only $e_g$ orbitals on a square lattice
with nearest-neighbor hopping,
it is useful for capturing the band topology of the $e_g$ orbitals.
It is also worth noting that
such simple tight-binding models,
including a model for graphene with Dirac points~\cite{Fujita1996},
the Kane--Mele model for topological insulators~\cite{Kane2005PRL95.146802},
and the Rashba--Hubbard model for Weyl semimetals~\cite{Kubo2024JPSJ},
have been instrumental in exploring topological phenomena.

Understanding the $e_g$ orbital system is crucial
not only because it serves as a prototypical multiorbital system
but also due to its relevance in various transition metal compounds.
These include manganese oxides~\cite{Jonker1950, vanSanten1950},
cuprate superconductors~\cite{Bednorz1986},
and the recently discovered
nickelate superconductor La$_3$Ni$_2$O$_7$~\cite{Sun2023}.
Therefore, a deeper understanding of the fundamental physics
of the $e_g$ orbital system can advance
research across a wide range of materials.

In our previous study,
we highlighted the importance of the band topology
of $e_g$ orbitals on a square lattice~\cite{Kubo2024PRB}.
A natural question arises:
What is the band topology of $e_g$ electrons in three-dimensional systems,
and what are its implications?
This question is particularly relevant
given the existence of several three-dimensional materials
with significant $e_g$ orbital contributions,
such as perovskite manganites~\cite{Jonker1950, vanSanten1950}.

Here, we extend the $e_g$ orbital model to a three-dimensional system,
specifically on a simple cubic lattice.
We find topological nodal lines in this model.
In the presence of such topological nodal lines,
surface states are expected~\cite{Heikkila2011, Burkov2011, Weng2015}.
For example, in a single-orbital tight-binding model on a diamond lattice,
a flat surface state appears on the (111) surface~\cite{Takagi2008, Takahashi2013}.
In our model, we find flat surface states
in the entire Brillouin zone for the (111) surface
as long as the bulk band projected onto the surface Brillouin zone is gapped.
This contrasts with the single-orbital model on a diamond lattice,
where the flat surface state appears
only in a part of the surface Brillouin zone.
The surface state in our model exhibits an octupole moment,
a characteristic feature of the $e_g$ orbital system.

\section{Model}
In this paper, we omit considerations of a magnetic field and magnetic ordering.
Consequently, we can disregard the spin degrees of freedom and reduce our model
to a spinless one.
The Hamiltonian for a simple cubic lattice with nearest-neighbor hopping
is given by
\begin{equation}
  H = \sum_{\tau \tau'}
  c_{\mib{k} \tau}^{\dagger}\epsilon_{\tau \tau'}(\mib{k}) c_{\mib{k} \tau'},
\end{equation}
where $c_{\mib{k} \tau}$ is the annihilation operator
for an electron with momentum $\mib{k}$ and orbital $\tau$.
Here, $\tau=1$ corresponds to the $x^2-y^2$ orbital
and $\tau=2$ corresponds to the $3z^2-r^2$ orbital.
The matrix elements of the Hamiltonian
are given by~\cite{Kubo2008JPSJ, Kubo2008JOAM}
\begin{align}
  \epsilon_{1 1}(\mib{k})
  &=
  \frac{1}{2}[3(dd\sigma)+(dd\delta)](c_x+c_y)
  +2(dd\delta)c_z,\\
  \epsilon_{2 2}(\mib{k})
  &=
  \frac{1}{2}[(dd\sigma)+3(dd\delta)](c_x+c_y)
  +2(dd\sigma)c_z,\\
  \epsilon_{1 2}(\mib{k})
  &=
  \epsilon_{2 1}(\mib{k})
  =
  -\frac{\sqrt{3}}{2}[(dd\sigma)-(dd\delta)](c_x-c_y),
\end{align}
where $c_{\mu} = \cos k_{\mu}a$ with $\mu=x$, $y$, or $z$,
and $a$ is the lattice constant.
$(dd\sigma)$ and $(dd\delta)$ represent
the two-center integrals~\cite{Slater1954}.
This model can be adapted to $\Gamma_8$ orbitals of $f$ electrons
by substituting $(dd\sigma)$ with $[3(ff\sigma)+4(ff\pi)]/7$
and $(dd\delta)$ with $[(ff\pi)+5(ff\delta)+15(ff\phi)]/21$.
Since the model is for two orbitals,
the matrix $\epsilon(\mib{k})$ can be expressed using the Pauli matrices:
\begin{equation}
  \begin{split}
    \epsilon(\mib{k})
    &=
    \frac{1}{2}[\epsilon_{11}(\mib{k})+\epsilon_{22}(\mib{k})]\sigma^0\\
    &\quad
    +\frac{1}{2}[\epsilon_{11}(\mib{k})-\epsilon_{22}(\mib{k})]\sigma^z
    +\epsilon_{12}(\mib{k})\sigma^x\\
    &=
    h_0(\mib{k})\sigma^0
    +h_x(\mib{k})\sigma^x
    +h_z(\mib{k})\sigma^z,
  \end{split}
\end{equation}
where $\sigma^{\mu}$ denotes the $\mu$ component of the Pauli matrix,
and $\sigma^0$ is the identity matrix.
The coefficients are given by:
\begin{align}
  \begin{split}
    h_0(\mib{k})
    &=
    [\epsilon_{11}(\mib{k})+\epsilon_{22}(\mib{k})]/2\\
    &= [(dd\sigma)+(dd\delta)](c_x+c_y+c_z)\\
    &= 2t_1(c_x+c_y+c_z),
  \end{split}\\
  \begin{split}
    h_x(\mib{k})
    &=
    \epsilon_{12}(\mib{k})
    =
    \epsilon_{21}(\mib{k})\\
    &=
    -\frac{\sqrt{3}}{2}[(dd\sigma)-(dd\delta)](c_x-c_y)\\
    &=
    -\sqrt{3}t_2(c_x-c_y),
  \end{split}\\
  \begin{split}
    h_z(\mib{k})
    &=
    [\epsilon_{11}(\mib{k})-\epsilon_{22}(\mib{k})]/2\\
    &= \frac{1}{2}[(dd\sigma)-(dd\delta)](c_x+c_y-2c_z)\\
    &= t_2(c_x+c_y-2c_z).
  \end{split}
\end{align}
We define the parameters:
\begin{align}
   t_1  &= [(dd\sigma)+(dd\delta)]/2,\\
   t_2 &= [(dd\sigma)-(dd\delta)]/2,
\end{align}
and assume $t_1 \ge 0$ and $t_2 \ge 0$ without loss of generality.
Thus, we parametrize them as
\begin{align}
  t_1  &= t \cos \alpha,\\
  t_2 &= t \sin \alpha,
\end{align}
with $t>0$ and $0 \le \alpha \le \pi/2$.

\begin{figure}
  \includegraphics[width=0.99\linewidth]
    {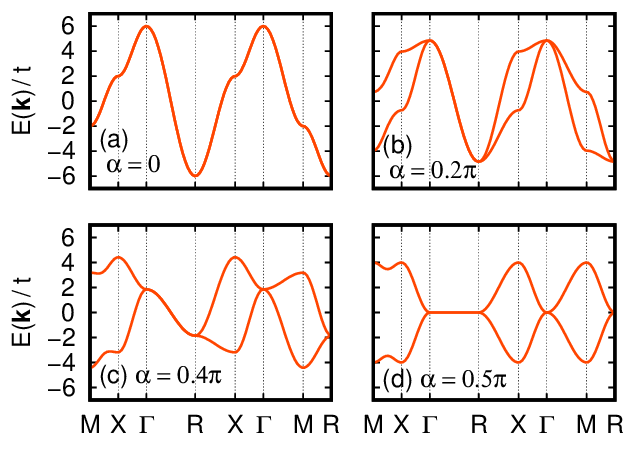}%
  \caption{
    (Color online)
    Energy dispersion along symmetric directions
    for (a) $\alpha=0$, (b) $\alpha=0.2\pi$, (c) $\alpha=0.4\pi$,
    and (d) $\alpha=0.5\pi$.
    The high-symmetry points are
    $\Gamma=(0,0,0)$,
    $X=(\pi/a,0,0)$,
    $M=(\pi/a,\pi/a,0)$, and
    $R=(\pi/a,\pi/a,\pi/a)$.
    \label{dispersion}}
\end{figure}

The model under consideration has been explored in diverse settings
for various values of the parameter $\alpha$.
In it on a two-dimensional lattice
with $\alpha = 0$
[$t_2 = 0$, $(dd\delta) = (dd\sigma)$; see Fig.\ref{dispersion}(a)],
prior studies have focused on
the potential for anisotropic superconducting pairing
due to the orbital anisotropy inherent to $e_g$ orbitals~\cite{Kubo2007PRB}.
Similar forms of anisotropic superconductivity have also been proposed
for $f$-electron systems, such as Pr$T_2X_{20}$
($T$ being a transition metal and $X$ representing Zn or Al)~\cite{Kubo2018JPSJ, Kubo2018AIPA, Kubo2020PRB, Kubo2020JPSCP}.
When $\alpha = \pi/4$
[$t_1 = t_2$, $(dd\delta) = 0$],
this model has been applied in two and three dimensions
to analyze physical properties in systems like
perovskite manganites~\cite{Anderson1959}
and $\Gamma_8$ $f$-electron orbitals,
where only $(ff\sigma)$ contributions are considered~\cite{Hotta2003,
Kubo2005PRB72.144401, Kubo2006JPSJ, Kubo2006PhysicaB, Kubo2006JPSJS,
Kubo2007JMMM, Kubo2017, Kubo2018JPCS, Kubo2018JPSJ, Kubo2018AIPA,
Kubo2020PRB, Kubo2020JPSCP}.
In the strong coupling regime,
an effective Hamiltonian derived from this model
highlights the orbital anisotropy-induced frustration~\cite{Ishihara1996, Ishihara1997, Brink1999, Kubo2002JPSJ71.1308, Kubo2002JPCS}.
Notably, the resulting frustrated behavior bears a resemblance
to the Kitaev model, which has been extensively investigated
in recent literature~\cite{Kitaev2006, Jackeli2009}.
For $\alpha = \pi/2$
[$t_1 = 0$, $(dd\delta) = -(dd\sigma)$; see Fig.\ref{dispersion}(d)],
the two-dimensional version of the model near half-filling exhibits
pocket Fermi surfaces located at $(\pm \pi/2, \pm \pi/2)$.
If superconducting pairs are formed by electrons
occupying the same Fermi pocket,
the resulting pairs acquire a finite total momentum of $(\pi, \pi)$.
This scenario is reminiscent of the Fulde--Ferrell--Larkin--Ovchinnikov
state~\cite{Fulde1964, Larkin1965},
but occurs here without an external magnetic field.
Such behavior parallels the $\eta$-pairing state
described in earlier works~\cite{Yang1989} and
has been further explored
in Refs.~\citen{Kubo2008JPSJ} and~\citen{Kubo2008JOAM}.

The energy dispersion of the model is given by
\begin{equation}
  \begin{split}
    E(\mib{k})
    &= h_0(\mib{k})\pm\sqrt{h_x^2(\mib{k})+h_z^2(\mib{k})}\\
    &= h_0(\mib{k})\pm h(\mib{k}).
  \end{split}
\end{equation}
Note that hybridization between the $x^2-y^2$ and $3z^2-r^2$ orbitals
is prohibited on the $k_x=k_y$ plane
due to mirror symmetry~\cite{Bishop2016, Horio2018}.
Additionally, these orbitals are degenerate along the $k_x=k_y=k_z$ line.
Consequently, the two bands intersect along this and equivalent lines,
$(k_x,k_y,k_z) \parallel (\pm 1, \pm 1, \pm 1)$.
These lines are topological nodal lines.

In Fig.~\ref{dispersion},
we show the energy dispersion for some values of $\alpha$.
The two bands are degenerate along the nodal line ($\Gamma$--$R$ line).
The bandwidth $W$ varies non-monotonically with $\alpha$:
$W=12t$ at $\alpha=0$ [Fig.~\ref{dispersion}(a)],
$W=6\sqrt{2}t$ at $\alpha=0.25\pi$,
$W=4\sqrt{5}t$ at $\alpha=\tan^{-1}2=0.3524\pi$,
and
$W=8t$ at $\alpha=0.5\pi$ [Fig.~\ref{dispersion}(d)].
The system is metallic except in the vicinity of $\alpha=0.5\pi$
around half-filling.
Around $\alpha=0.5\pi$ and close to half-filling,
it is semimetallic with the nodal lines.

We can define the winding number $w$ of
the normalized two-component vector field
$\hat{\mib{h}}(\mib{k})=\mib{h}(\mib{k})/h(\mib{k})
=[h_x(\mib{k}),h_z(\mib{k})]/h(\mib{k})
=[\hat{h}_x(\mib{k}),\hat{h}_z(\mib{k})]$ as follows,
\begin{equation}
  w
  =
  \oint_{C} \frac{d\mib{k}}{2\pi}
  \cdot
  \left[
    \hat{h}_x(\mib{k})\mib{\nabla}\hat{h}_z(\mib{k})
    -\hat{h}_z(\mib{k})\mib{\nabla}\hat{h}_x(\mib{k})\right],
\end{equation}
where $C$ is a closed loop.
For a loop encircling
the nodal line $(0,0,0)$--$(a_x \pi/a, a_y \pi/a, a_z\pi/a)$
with $a_x$, $a_y$, $a_z=\pm 1$,
we obtain $w=-a_x a_y a_z$.

In a model with finite thickness, i.e.,
with two parallel surfaces,
we define the momentum $\mib{k}_{\parallel}$ parallel to the surfaces.
When $\mib{k}_{\parallel}$ is fixed at a certain value,
the system can be regarded as a one-dimensional system
along the direction perpendicular to the surfaces.
We denote the lattice constant of this one-dimensional model as $\tilde{a}$.
The momentum $k_{\perp}$ of this one-dimensional system
is defined in the limit of infinite thickness.
The winding number $w(\mib{k}_{\parallel})$
for the one-dimensional system is then given by
\begin{equation}
  w(\mib{k}_{\parallel})
  =
  \int_0^{2\pi/\tilde{a}} \frac{dk_{\perp}}{2\pi}
  \left[
  \hat{h}_x(\mib{k}) \frac{\partial}{\partial k_{\perp}} \hat{h}_z(\mib{k})
  -\hat{h}_z(\mib{k}) \frac{\partial}{\partial k_{\perp}} \hat{h}_x(\mib{k})
  \right].
\end{equation}
If this one-dimensional system with a fixed $\mib{k}_{\parallel}$
has a band gap with a nonzero winding number $w(\mib{k}_{\parallel})$,
it can be regarded as a one-dimensional topological insulator
and will exhibit edge states.
These edge states correspond to surface states in the original system
on a simple cubic lattice at $\mib{k}_{\parallel}$.
Therefore, surface states appear at $\mib{k}_{\parallel}$
for a nonzero $w(\mib{k}_{\parallel})$,
provided that the bulk bands projected onto the surface Brillouin zone
have a gap at this point.
Since the surface states are topologically protected,
we expect them to be robust to some extent
against perturbations such as disorder.
However, a magnetic field or magnetic impurities might alter
the situation because they break the time-reversal symmetry of the model.

Since $E_g \times E_g = A_{1g} + A_{2u} + E_g$,
the multipole operators in the $e_g$ orbitals are
the charge operator $Q$ ($A_{1g}$),
the octupole operator $T_{xyz}$ ($A_{2u}$),
and the quadrupole operators $O_{20}$ and $O_{22}$ ($E_g$).
These operators are expressed in terms of
the orbital angular momentum operator $\mib{l}$
and can be represented using Pauli matrices in the $e_g$ orbital basis
as follows~\cite{Takahashi2000, Kubo2002JPSJ71.183}:
$Q \propto 1 \propto \sigma^0$,
$T_{xyz} \propto \{ l_x l_y l_z \} \propto \sigma^y$,
$O_{20} \propto 3l_z^2 - l(l+1) \propto \sigma^z$,
and $O_{22} \propto l_x^2 - l_y^2 \propto \sigma^x$,
where $\{ \cdots \}$ denotes the symmetrized product.
Since the coefficients of the multipole operators are not critical
for the present study, they are set to unity.

\begin{figure}
  \begin{center}
    \includegraphics[width=0.82\linewidth]
    {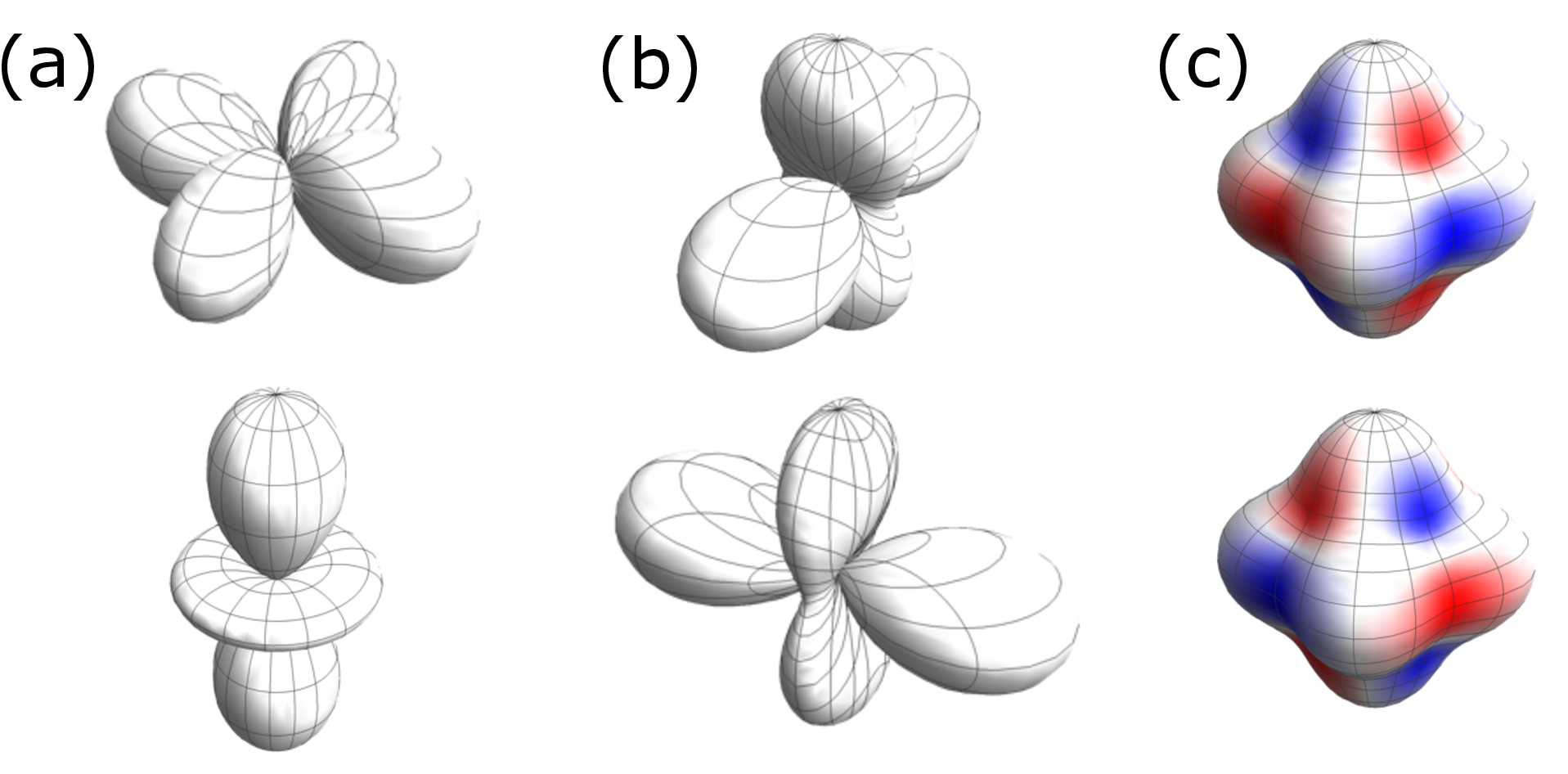}%
  \end{center}
  \caption{
    (Color online)
    Schematic view of the eigenstates of the multipole operators:
    (a) $O_{20}$, (b) $O_{22}$, and (c) $T_{xyz}$.
    The shapes illustrate
    the charge distributions in the eigenstates,
    with red (blue) indicating the dipole moment density
    parallel (antiparallel) to the $z$ direction.
    \label{multipoles}}
\end{figure}

Within the $e_g$ orbitals, in the absence of spin degrees of freedom,
the matrix elements of the dipole moments are zero.
However, the octupole moment describes an anisotropic distribution
of the dipole moments,
similar to a quadrupole moment describing
an anisotropic charge distribution (see Fig.~\ref{multipoles}).
Typically, such a higher-order multipole is classified into the same symmetry
of a lower-order multipole, such as dipoles or quadrupoles,
when considering a crystalline electric field.
In such a situation, higher-order multipoles do not emerge
as independent degrees of freedom,
making it sufficient to consider only the lower-order multipoles.
In the case of $e_g$ orbitals, however,
the octupole moment is independent of the lower-order multipoles,
and thus, it has the potential to influence physical properties.

The multipole operators at site $(i,\mib{r}_{\parallel})$ are given as
\begin{equation}
  \hat{\sigma}^{\mu}(i,\mib{r}_{\parallel})
  =
  \sum_{\tau \tau'}c^{\dagger}_{i \mib{r}_{\parallel} \tau}
  \sigma^{\mu}_{\tau \tau'} c_{i \mib{r}_{\parallel} \tau'},  
\end{equation}
where $\mib{r}_{\parallel}$ is the position vector parallel to the surfaces,
$i$ denotes the position perpendicular to the surfaces,
$c_{i \mib{r}_{\parallel} \tau}$ is the annihilation operator
of the electron at site $(i,\mib{r}_{\parallel})$ with orbital $\tau$,
and $\mu=0$, $x$, $y$, or $z$.
The multipole operator at position $i$ is given by
\begin{equation}
  \begin{split}
    \hat{\sigma}^{\mu}(i)
    &=
    \sum_{\mib{r}_{\parallel}} \hat{\sigma}^{\mu}(i,\mib{r}_{\parallel})\\
    &=
    \sum_{\mib{k}_{\parallel} \tau \tau'}
    c^{\dagger}_{i \mib{k}_{\parallel} \tau}
    \sigma^{\mu}_{\tau \tau'} c_{i \mib{k}_{\parallel} \tau'}\\
    &=
    \sum_{\mib{k}_{\parallel}} \hat{\sigma}^{\mu}(i,\mib{k}_{\parallel}),
  \end{split}
\end{equation}
where $c_{i \mib{k}_{\parallel} \tau}$ is the Fourier transform of
$c_{i \mib{r}_{\parallel} \tau}$ parallel to the surfaces.
The zeroth component is the charge operator $\hat{Q}(i)=\hat{\sigma}^0(i)$,
the $z$ and $x$ components are the quadrupole operators
$\hat{O}_{20}(i)=\hat{\sigma}^z(i)$ and $\hat{O}_{22}(i)=\hat{\sigma}^x(i)$,
respectively,
and the $y$ component is the octupole operator
$\hat{T}_{xyz}(i)=\hat{\sigma}^y(i)$.

Since the octupole moment is odd under time reversal,
we expect the octupole moments at $\mib{k}_{\parallel}$
and $-\mib{k}_{\parallel}$ to have opposite signs with the same magnitude
due to the time-reversal symmetry of the model.
Additionally, a surface state at $\mib{k}_{\parallel}$ is equivalent to
a state on the opposite surface at $-\mib{k}_{\parallel}$.
Thus, we anticipate opposite signs of the octupole moments
with the same magnitude
on opposite surfaces for a given finite $\mib{k}_{\parallel}$.
In other words, it is natural to expect
that finite octupole moments appear in such pairs in the surface states.

\section{(001) surface}
For the (001) surface case,
$\mib{k}_{\parallel}=(k_x,k_y)$ and $k_{\perp}=k_z$.
The lattice constant in the direction
perpendicular to the surfaces is $\tilde{a}=a$.
We find $w(\mib{k}_{\parallel})=0$ in the entire surface Brillouin zone.

\begin{figure}
  \includegraphics[width=0.99\linewidth]
    {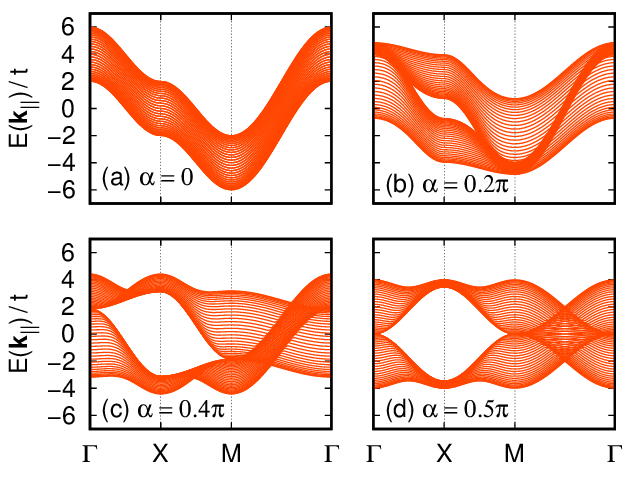}%
  \caption{
    (Color online)
    Energy dispersion for a system with (001) surfaces
    along symmetric directions
    for (a) $\alpha=0$, (b) $\alpha=0.2\pi$,
    (c) $\alpha=0.4\pi$, and (d) $\alpha=0.5\pi$.
    The high-symmetry points are
    $\Gamma=(0,0)$,
    $X=(\pi/a,0)$, and
    $M=(\pi/a,\pi/a)$.
    The number of layers parallel to the surfaces is $L_z=31$.
    \label{band_001edge}}
\end{figure}

In Fig.~\ref{band_001edge},
we show the energy dispersion $E(\mib{k}_{\parallel})$
for a lattice with a finite thickness of
$\tilde{a} L_z$, where $L_z=31$.
As expected from $w(\mib{k}_{\parallel})=0$,
we observe no surface states isolated from the bulk states.

\section{(110) surface}
For the (110) surface case,
the lattice constant in the direction
perpendicular to the surfaces is $\tilde{a}=a/\sqrt{2}$.
We define $\mib{k}_{\parallel}=(k_y',k_z)$ and $k_{\perp}=k_x'$,
where $k_y'$ is the momentum parallel to the surfaces
and perpendicular to the $z$ axis,
and $k_x'$ is the momentum perpendicular to the surfaces.
The vector $\mib{h}(\mib{k})$ is expressed by using these momenta:
$h_x(\mib{k})=-2\sqrt{3}t_2\sin(k_x'a/\sqrt{2})\sin(k_y'a/\sqrt{2})$
and
$h_z(\mib{k})=2t_2[\cos(k_x'a/\sqrt{2})\cos(k_y'a/\sqrt{2})-c_z]$.

\begin{figure}
  \begin{center}
    \includegraphics[width=0.6\linewidth]
    {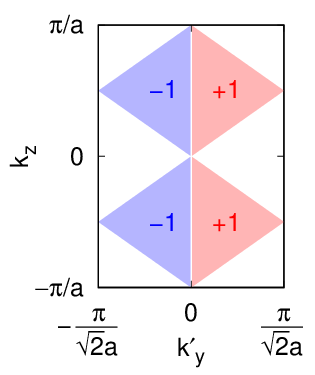}%
  \end{center}
  \caption{
    (Color online)
    Winding number $w(\mib{k}_{\parallel})$
    in the first Brillouin zone
    for a system with (110) surfaces.
    The winding number $w(\mib{k}_{\parallel})$ is zero in the non-shaded areas.
    \label{winding_number_110edge}}
\end{figure}

In Fig.~\ref{winding_number_110edge},
we show the winding number $w(\mib{k}_{\parallel})$
in the surface Brillouin zone.
In the region where $w(\mib{k}_{\parallel}) \ne 0$,
we expect a surface state,
provided that the bulk band projected onto the surface Brillouin zone is gapped.

\begin{figure}
  \includegraphics[width=0.99\linewidth]
    {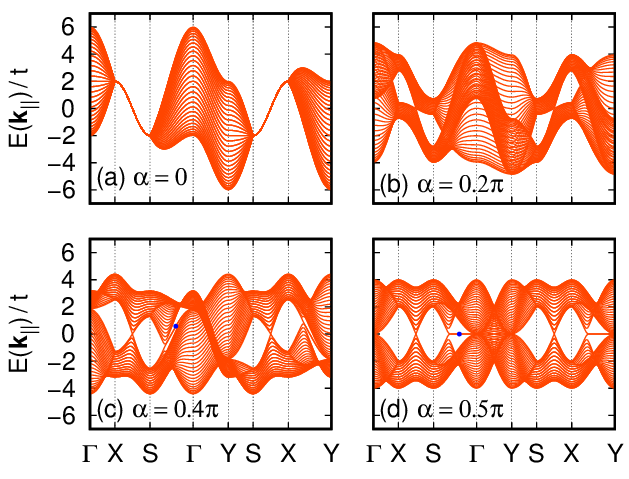}%
  \caption{
    (Color online)
    Energy dispersion for a system with (110) surfaces
    along symmetric directions
    at (a) $\alpha=0$, (b) $\alpha=0.2\pi$,
    (c) $\alpha=0.4\pi$, and (d) $\alpha=0.5\pi$.
    The high-symmetry points are
    $\Gamma=(0,0)$,
    $X=(\pi/\sqrt{2}a,0)$, 
    $Y=(0,\pi/a)$, and
    $S=(\pi/\sqrt{2}a,\pi/a)$.
    The number of layers parallel to the surfaces is $L_{x'}=31$.
    The solid circles in (c) and (d) indicate the surface states
    in which the multipole density is evaluated
    in Figs.~\ref{multipole_110edge}(a) and \ref{multipole_110edge}(b),
    respectively.
    \label{band_110edge}}
\end{figure}

In Fig.~\ref{band_110edge},
we show the energy dispersion $E(\mib{k}_{\parallel})$
for a lattice with a finite thickness $\tilde{a} L_{x'}$
with $L_{x'}=31$.
For $\alpha=0$, all the bands are degenerate for $k_y'=\pi/\sqrt{2}a$.
For $\alpha < \tan^{-1}2=0.3524\pi$,
the bulk band gap is closed in the region where $w(\mib{k}_{\parallel}) \ne 0$.
For $\alpha > \tan^{-1}2$,
the bulk band gap opens except on the lines
where $w(\mib{k}_{\parallel})$ changes.
For $\alpha > \tan^{-1}2$,
we observe surface states isolated from the bulk band
in parts of $S$--$\Gamma$ and $X$--$Y$ lines,
as shown in Figs.~\ref{band_110edge}(c) and \ref{band_110edge}(d).
The appearance of the surface states is in accord with
the $w(\mib{k}_{\parallel}) \ne 0$ region in Fig.~\ref{winding_number_110edge}.

For a fixed $k_z$,
the model is reduced to a two-dimensional model
with the difference in the energy levels of orbitals $\Delta=-4t_2 c_z$
and an energy shift of $2t_1 c_z$.
Using the results of the two-dimensional model~\cite{Kubo2024PRB},
we obtain the energy of the surface state:
$E = -\Delta/\tan\alpha+2t_1c_z = 6t_1 c_z$.
The energy dispersions of the surface state
on the $S$--$\Gamma$ and $X$--$Y$ lines
for $\alpha=0.4\pi$ in Fig.~\ref{band_110edge}(c)
and
for $\alpha=0.5\pi$ in Fig.~\ref{band_110edge}(d)
are given by this equation.

From the results of the two-dimensional model,
we also conclude that,
at $(k_y',k_z)=[\pi/(3\sqrt{2}a),\pi/(2a)]$,
the surface state is completely localized
on the first layer of the surface
with a fully polarized octupole moment for $\alpha=0.5\pi$.
For other $\mib{k}_{\parallel}$ and $\alpha$,
we need to evaluate the multipole density numerically.

\begin{figure}
  \begin{center}
    \includegraphics[width=0.8\linewidth]
    {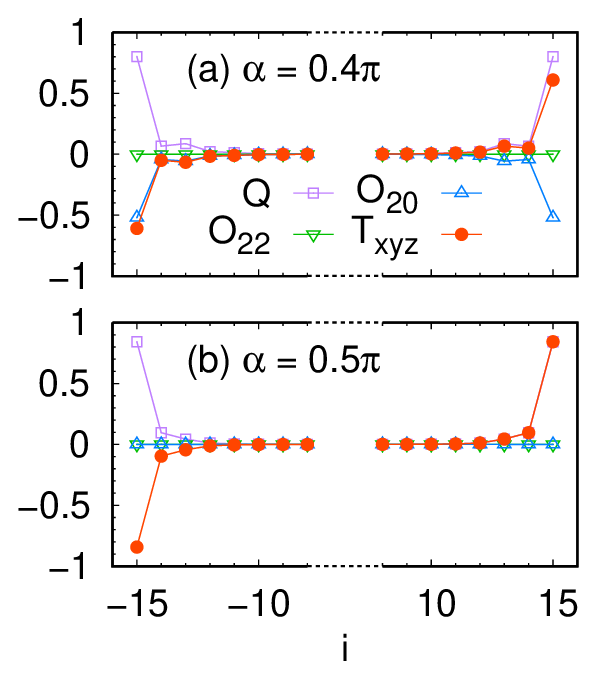}%
  \end{center}
  \caption{
    (Color online)
    Multipole density in the surface state
    for a lattice with (110) surfaces
    at $\mib{k}_{\parallel}=0.4 \times S$
    (a) for $\alpha=0.4\pi$
    [see Fig.~\ref{band_110edge}(c)]
    and
    (b) for $\alpha=0.5\pi$
    [see Fig.~\ref{band_110edge}(d)].
    The number of layers parallel to the surfaces is $L_{x'}=31$,
    and $S=(\pi/\sqrt{2}a,\pi/a)$.
    \label{multipole_110edge}}
\end{figure}

We show the multipole density around the surfaces
for $\alpha=0.4\pi$ in Fig.~\ref{multipole_110edge}(a) and
for $\alpha=0.5\pi$ in Fig.~\ref{multipole_110edge}(b)
at $\mib{k}_{\parallel}=0.4 \times S$
with $S=(\pi/\sqrt{2}a,\pi/a)$.
The surface states are doubly degenerate,
and we show the total multipole density of these two states.
From the charge density $Q(i)=\langle \hat{Q}(i) \rangle$,
where $\langle \cdots \rangle$ denotes the sum of the expectation values
in the two surface states,
we recognize that these states are localized around the surfaces.
$O_{22}(i)=\langle \hat{Q}_{22}(i) \rangle$ is always zero,
and $O_{20}(i)=\langle \hat{Q}_{20}(i) \rangle$ is also zero for $\alpha=0.5\pi$.
The octupole moment $T_{xyz}(i)=\langle \hat{T}_{xyz}(i) \rangle$
has opposite signs between the surfaces.

\begin{figure}
  \includegraphics[width=0.99\linewidth]
    {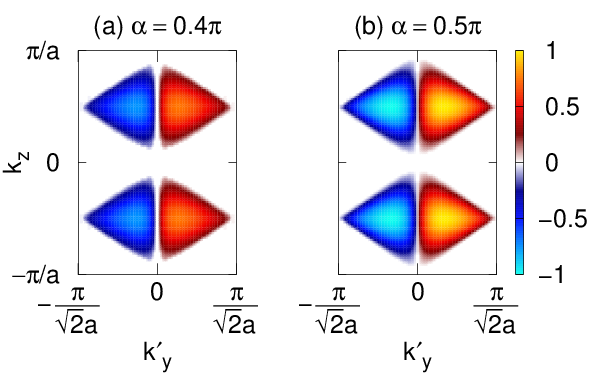}%
  \caption{
    (Color online)
    Octupole density on the top layer $i=(L_{x'}-1)/2$
    of the (110) surface in the surface state
    (a) for $\alpha=0.4\pi$ 
    and
    (b) for $\alpha=0.5\pi$.
    The number of layers parallel to the surfaces is $L_{x'}=101$.
    \label{octupole_110edge}}
\end{figure}

In Fig.~\ref{octupole_110edge},
we show the octupole density on the top layer $i=(L_{x'}-1)/2$
in the surface state.
As we noted, at $(k_y',k_z)=[\pi/(3\sqrt{2}a),\pi/(2a)]$
(at the center of mass of the triangle)
for $\alpha=0.5\pi$,
the octupole density reaches unity.

\section{(111) surface}
For the (111) surface case,
the lattice constant in the direction
perpendicular to the surfaces is $\tilde{a}=a/\sqrt{3}$.
We define $\mib{k}_{\parallel}=(k_x',k_y')$ and $k_{\perp}=k_z'$.
The top layer of the (111) surface forms a triangular lattice.
Here, $k_x'$ is the momentum parallel to one of the edges of a triangle
in the triangular lattice,
$k_y'$ is the momentum parallel to the surfaces
and perpendicular to the $k_x'$ axis,
and $k_z'$ is the momentum perpendicular to the surfaces.
The vector $\mib{h}(\mib{k})$ is expressed using these momenta as follows:
$h_x(\mib{k})
=2\sqrt{3}t_2\sin(k_x'a/\sqrt{2})\sin(k_y'a/\sqrt{6}-k_z'a/\sqrt{3})$
and
$h_z(\mib{k})
=2t_2[\cos(k_x'a/\sqrt{2})\cos(k_y'a/\sqrt{6}-k_z'a/\sqrt{3})
-\cos(2k_y'a/\sqrt{6}+k_z'a/\sqrt{3})]$.
The coefficient of the unit matrix is
$h_0(\mib{k})=2t_1
[\cos(k_x'a/\sqrt{2}+k_y'a/\sqrt{6}-k_z'a/\sqrt{3})
+\cos(k_x'a/\sqrt{2}-k_y'a/\sqrt{6}+k_z'a/\sqrt{3})
+\cos(2k_y'a/\sqrt{6}+k_z'a/\sqrt{3})]$.
Since $\epsilon(\mib{k}) \rightarrow -\epsilon(\mib{k})$
under $k_z' \rightarrow k_z'+\pi/\tilde{a}$,
the bulk energy spectrum projected onto the (111) surface $E(\mib{k}_{\parallel})$
(collected over $k_z'$) is symmetric with respect to $E(\mib{k}_{\parallel})=0$.

\begin{figure}
  \begin{center}
    \includegraphics[width=0.8\linewidth]
    {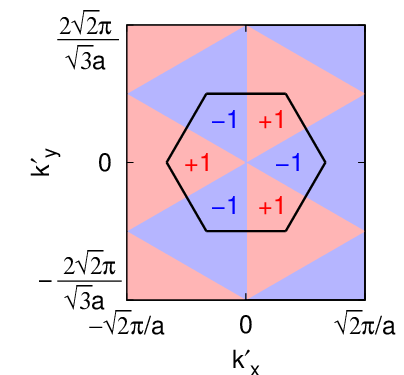}%
  \end{center}
  \caption{
    (Color online)
    Winding number $w(\mib{k}_{\parallel})$
    for a system with (111) surfaces.
    The area enclosed by the solid line is the first Brillouin zone.
    \label{winding_number_111edge}}
\end{figure}

In Fig.~\ref{winding_number_111edge},
we show the winding number $w(\mib{k}_{\parallel})$
in the reciprocal space parallel to the surfaces.
The winding number is nonzero except along the $\Gamma$--$M$
[(0,0)--$(0,\sqrt{2}\pi/\sqrt{3}a)$] and equivalent lines.
Thus, we expect a surface state at any $\mib{k}_{\parallel}$
as long as the bulk band projected onto the surface Brillouin zone is gapped
there.

\begin{figure}
  \includegraphics[width=0.99\linewidth]
    {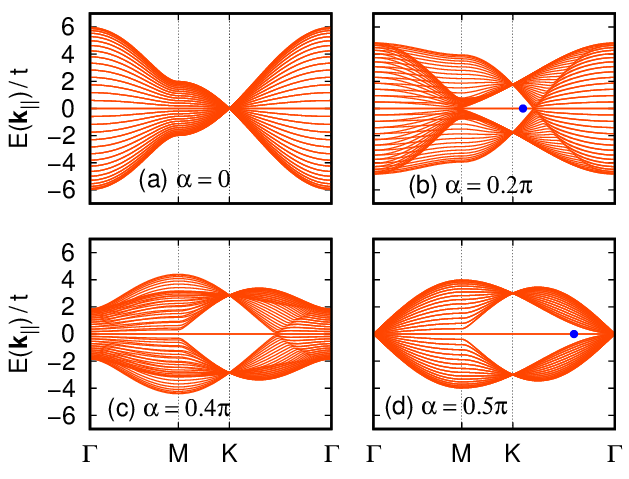}%
  \caption{
    (Color online)
    Energy dispersion for a system with (111) surfaces
    along symmetric directions
    at (a) $\alpha=0$, (b) $\alpha=0.2\pi$,
    (c) $\alpha=0.4\pi$, and (d) $\alpha=0.5\pi$.
    The high-symmetry points are
    $\Gamma=(0,0)$,
    $M=(0,\sqrt{2}\pi/\sqrt{3}a)$, and
    $K=(\sqrt{2}\pi/3a,\sqrt{2}\pi/\sqrt{3}a)$.
    The number of layers parallel to the surfaces is $L_{z'}=31$.
    The solid circles in (b) and (d) indicate the surface states
    in which the multipole density is evaluated
    in Figs.~\ref{multipole_111edge}(a) and \ref{multipole_111edge}(b),
    respectively.
    \label{band_111edge}}
\end{figure}

In Fig.~\ref{band_111edge},
we show the energy dispersion $E(\mib{k}_{\parallel})$
for a lattice with a finite thickness $\tilde{a} L_{z'}$
with $L_{z'}=31$.
The size effect is large around the $M$ point for even numbers of $L_{z'}$;
therefore, we have chosen $L_{z'}=31$.
We observe the flat surface state isolated from the bulk band
when the bulk band gap opens.
In particular, for $\alpha=0.5\pi$,
the flat surface state extends in the entire Brillouin zone
except on the lines where $w(\mib{k}_{\parallel})$ changes.

We can show that
the surface state is completely localized
on the first layer of the surface with a fully polarized octupole moment
at the $K$ point,
irrespective of the value of $\alpha$.
At other $\mib{k}_{\parallel}$,
we need to evaluate the multipole density numerically.

\begin{figure}
  \begin{center}
    \includegraphics[width=0.8\linewidth]
    {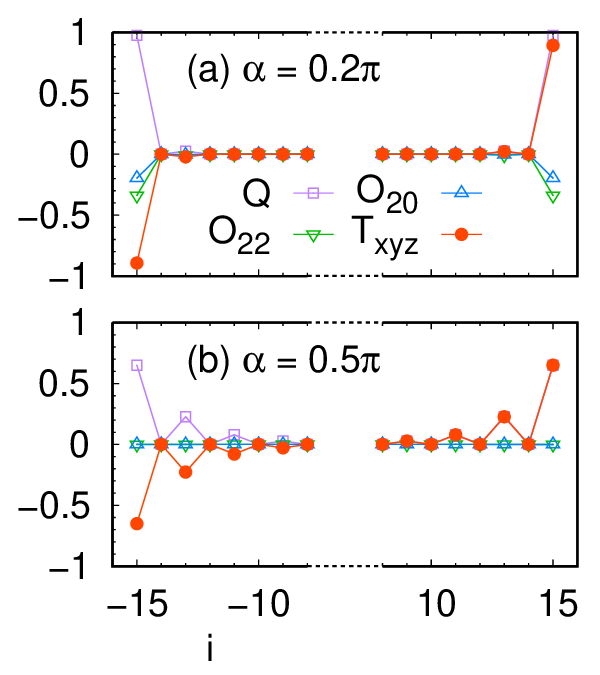}%
  \end{center}
  \caption{
    (Color online)
    Multipole density in the surface state for a system with (111) surfaces
    (a) for $\alpha=0.2\pi$ at $\mib{k}_{\parallel}=0.9 \times K$
    [see Fig.~\ref{band_111edge}(b)]
    and
    (b) for $\alpha=0.5\pi$ at $\mib{k}_{\parallel}=0.4 \times K$
    [see Fig.~\ref{band_111edge}(d)].
    The number of layers parallel to the surfaces is $L_{z'}=31$,
    and $K=(\sqrt{2}\pi/3a,\sqrt{2}\pi/\sqrt{3}a)$.
    \label{multipole_111edge}}
\end{figure}

We show the multipole density around the surfaces
for $\alpha=0.2\pi$ in Fig.~\ref{multipole_111edge}(a) and
for $\alpha=0.5\pi$ in Fig.~\ref{multipole_111edge}(b).
The surface states are doubly degenerate,
and we show the total multipole density of these two states.
From the charge density $Q(i)$,
we recognize that these states are localized around the surfaces.
$O_{20}(i)$ and $O_{22}(i)$ are zero for $\alpha=0.5\pi$.
The octupole moment $T_{xyz}(i)$ has opposite signs between the surfaces.

\begin{figure}
  \includegraphics[width=0.99\linewidth]
    {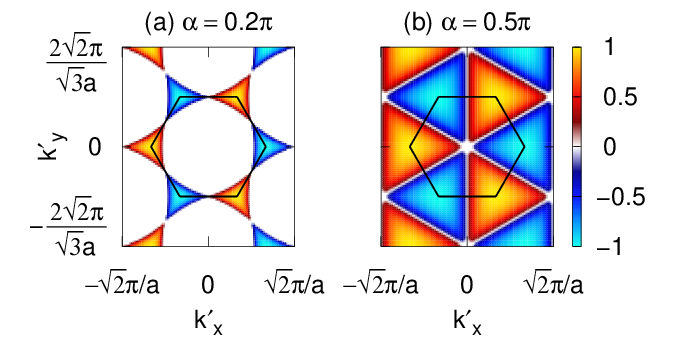}%
  \caption{
    (Color online)
    Octupole density on the top layer $i=(L_{z'}-1)/2$ of the (111) surface
    in the surface state
    (a) for $\alpha=0.2\pi$ 
    and
    (b) for $\alpha=0.5\pi$.
    The number of layers parallel to the surfaces is $L_{z'}=101$.
    \label{octupole_111edge}}
\end{figure}

In Fig.~\ref{octupole_111edge},
we show the octupole density on the top layer $i=(L_{z'}-1)/2$
in the surface state.
The octupole moment reaches unity at the $K$ point,
irrespective of the value of $\alpha$.

\section{Summary and Discussion}
We have studied the surface states
of an $e_g$ orbital model on a simple cubic lattice
with nearest-neighbor hopping.
Our findings reveal surface states with octupole moments
on the (110) and (111) surfaces.
Currently, it is unclear whether octupole moments
have a significant impact on experiments.
However, we believe that the identification of surface octupole moments
in a typical multiorbital model is noteworthy.

Concerning octupole moments,
ordering of them has been observed in $f$-electron materials,
such as NpO$_2$~\cite{Paixao2002, Caciuffo2003, Tokunaga2005, Kubo2005PRB71,
  Kubo2005PRB72.144401, Kubo2005PRB72.132411, Kubo2006PhysicaB, Kubo2006JPSJS}
and Ce$_x$La$_{1-x}$B$_6$~\cite{Kubo2003, Kubo2004, Morie2004, Mannix2005,
  Kusunose2005, Kuwahara2007}.
In $e_g$ orbital systems of $d$ electrons,
the possibility of octupole ordering in the bulk perovskite manganites
has been theoretically explored~\cite{Takahashi2000, Maezono2000, vandenBrink2001, Khomskii2001}.
However, a theoretical analysis that accounts for fluctuations
beyond the mean-field framework determined that
the emergence of such ordering is improbable~\cite{Kubo2002JPSJ71.183, Kubo2002JPSJS, Kubo2002JPCS}.

Although the ordering of octupole moments in $e_g$ orbitals may be challenging,
this study reveals the potential for octupole moments
to emerge on the surfaces of $e_g$ orbital systems.
However, detecting these surface octupole moments is difficult,
as they have opposite signs for opposite momenta,
leading to a net momentum-averaged value of zero.
Therefore, momentum-resolved experiments are required
to detect the surface octupole moments,
making this a challenging experimental task.
Nevertheless, detecting the surface state itself may still be possible
without distinguishing the octupole moment.

Another characteristic of the surface state is the region in which it appears.
The flat surface state appears on the (111) surface
across all surface momenta, provided the bulk band gap is open.
This contrasts with the surface state
of the single-orbital model on a diamond lattice with nearest-neighbor hopping,
where the surface state emerges
only in specific regions of the surface Brillouin zone~\cite{Takagi2008}.

Around a surface,
the symmetry is lowered, which can lift the degeneracy
of the $e_g$ orbitals~\cite{Fulde1973, Terakura1975}.
For the (001) and (110) surfaces,
the degeneracy is lifted,
and the surface state of the (110) surface can be influenced
by this effect.
In contrast, for the (111) surface,
the $e_g$ orbitals remain degenerate,
making the surface state robust against this symmetry-lowering effect.

In the presence of spin degrees of freedom and spin-orbit coupling,
magnetic dipole moments can also emerge in the surface states
alongside the magnetic octupole moments.
However, the effects of spin-orbit coupling primarily occur
through the $t_{2g}$ level.
Therefore, these effects are expected to be weak,
provided that the splitting between the $e_g$ and $t_{2g}$ levels
is sufficiently large, which is one of the assumptions in this study.

To realize the surface state in an actual material,
a system with nearly half-filled $e_g$ bands is desirable.
As a candidate material, BaFeO$_3$ is promising~\cite{Hayashi2011}.
This material crystallizes in the perovskite structure,
with the Fe$^{4+}$ ion possessing a $3d^4$ electron configuration.
The ground state is ferromagnetic.
Assuming the spins are fully polarized,
the present spinless model can be applied to the half-filled $e_g$ orbitals,
while the wholly filled $t_{2g}$ orbitals can be ignored.
Another potential candidate is a system
with ions that have a $3d^8$ electron configuration
in the paramagnetic state.
LaCuO$_3$, a Cu$^{3+}$ ($3d^8$) compound~\cite{Demazeau1972},
could serve as a candidate for the present model,
although electron correlations must be considered
to accurately describe this compound.

\begin{acknowledgment}
This work was supported by JSPS KAKENHI Grant No.
JP23K03330. 
\end{acknowledgment}


\end{document}